\documentstyle[aps,multicol,epsfig,prl]{revtex}

\def\kp{k_{y}}
\def\ko{k_{x}}

\def\be{\begin{equation}}
\def\ee{\end{equation}}
\def\ba{\begin{eqnarray}}
\def\ea{\end{eqnarray}}

\def\LSCO{La$_{2-x}$Sr$_x$CuO$_4$}

\def\YBCO{YBa$_2$Cu$_3$O$_{7-\delta}$}

\def\BSCCO{Bi$_2$Sr$_2$CaCu$_2$O$_{8+\delta}$}
\def\C60{A$_x$C$_{60}$}
\def\LNSCO{La$_{1.6-x}$Nd$_{0.4}$Sr$_x$CuO$_{4}$}

\input epsf

\begin{document}

\title
{Distribution of spectral weight in a system with disordered stripes}

\author{M.~Granath, V.~Oganesyan, D.~Orgad and S.~A.~Kivelson}
\address
{ Department of Physics, UCLA, Los Angeles, California  90095}

\date{\today}
\maketitle
\begin{abstract}
The ``band-structure'' of a disordered stripe array is
computed and compared, at a qualitative level, to angle resolved 
photoemission experiments on the cuprate high temperature
superconductors. The low-energy states are found to be
strongly localized transverse to the stripe direction, so
the electron dynamics is strictly one-dimensional (along
the stripe). Despite this, aspects of the two dimensional
band-structure Fermi surface are still vividly apparent.
\end{abstract}

\begin{multicols}{2}
\narrowtext
\section{introduction}
There is strong evidence, in part based on careful analysis of single
particle spectral functions deduced from angle resolved photoemission (ARPES),
that the normal state of the high temperature superconductors is not a 
conventional Fermi liquid.  At the same time, there is clear evidence that in
some, and possibly in all of these materials, there 
are local, self-organized quasi-one-dimensional structures, 
``stripes'', which substantially affect the electron dynamics. Many
striking features of the ARPES spectrum have a natural 
interpretation in terms of an electronic structure dominated by
stripes\cite{salkola,hanke,machida,lnsco,erica,dror}. 

One such feature, 
which is reproduced in 
Fig.~\ref{arpes}, is that the low-energy spectral weight lies predominantly 
in regions of the Brillouin zone (BZ) in the vicinity of the $(\pm\pi,0)$ 
and $(0,\pm\pi)$ points.  (These are often referred to as the 
``anti-nodal regions'' as this is where the maximum in the d-wave 
superconducting gap occurs.) These regions have very straight boundaries 
which run parallel to the momentum axes and are displaced by approximately
$\pm\pi/4$ from them. This spectral weight has been interpreted in terms 
of quarter filled stripes (so $k_F=\pi/4$) along the $a$ and $b$ 
directions of the lattice \cite{salkola,lnsco,dual}.

However, a number of features of the ARPES data are inconsistent with
the most naive picture of an array of strictly one-dimensional quarter 
filled stripes. Of these, the most obvious is the occurrence of features, 
especially in the ``nodal regions'' of the BZ [around $(\pm\pi/2,\pm\pi/2)$],
which are reminiscent of the underlying, two dimensional band structure.  
An example is the appearance of low-energy spectral weight along segments 
parallel to the lines connecting the anti-nodal points in \LSCO (LSCO) 
and \LNSCO (LNSCO). A similar signature  also appears in Pb-doped \BSCCO 
(BSCCO), as shown in Fig.~\ref{arpes}.

It is the principal purpose of this paper to elucidate how this apparently 
two-dimensional structure arises naturally from a slightly more sophisticated 
analysis of the band-structure of a stripy system. In doing so we extend 
the initial work of Ref. \ref{salkolaref} and complement a recent study 
of the effects of realistic band parameters on the spectrum by Fleck, 
Pavarini and Andersen \cite{realistic}. We show that a system 
can exhibit what seems to be a Fermi surface of a two-dimensional metal   
despite the fact that the {\em dynamics} of its low-energy electrons  
is entirely one-dimensional~!  
This conclusion holds even in the presence of interactions that preclude
any quasiparticle-like description of the system. Under such conditions 
although the low-energy features in momentum space are sharp, the spectral 
function, considered as function of energy for fixed momentum
can be broad\cite{dror,valla}.  
 
Our results are easily summarized.
We have computed the band-structure of electrons in the potential generated 
by a typical configuration of the ``slow'' collective fields that define the 
stripe order.  (The explicit Hamiltonian is presented in Section \ref{model}.)
Figs.~\ref{random} and \ref{fig418} depict the $\vec k$ space
distribution of low-energy spectral intensity in the first BZ for 
disordered and ordered arrays of parallel stripes, respectively. 
In both cases most of the low-energy spectral weight is concentrated in the 
anti-nodal regions. However, while the ordered stripe array exhibits a 
spectral gap in the nodal region, the disordered array has low-energy weight 
there, much like that seen in experiment. 

We picture the disordered array as 
being a ``snapshot'' of a ``fluctuating stripe array,'' although it 
could also reflect the effect of quenched disorder.
The low-energy electronic states that are responsible for its spectral map,
shown in Fig.~\ref{random}, lie within the Mott gap. They are bound states
that decay exponentially in the direction perpendicular 
to the stripes.  However, they are extended along the stripes, 
so the low energy electron dynamics is strictly
one-dimensional.  The resulting band-structure, which is shown in 
Fig.~\ref{randomdispfig}, is, to a good approximation, 
a superposition of the band-structures of single isolated stripes and of  
small clusters (typically pairs)  of proximate stripes. 
  
The band-structure and spectral distribution of a single stripe are shown 
in Figs.~\ref{singledispfig} and \ref{singleAfig}. Here, the fact 
that the electron wavefunctions in the gap have a non-negligible 
extent transverse to the stripe can be seen to produce an image of the 
full two-dimensional band structure. However, as is apparent from the 
spectrum, unless an isolated stripe is nearly half-filled (which is 
physically implausible), it will not have any low-energy spectral weight 
near ($\pi/2,\pi/2$). This is the reason for the absence of spectral weight 
in the nodal region of the ordered array since its wavefunctions are Bloch 
states constructed from single stripe bound states. 

In the disordered array of stripes, the spectral weight in the nodal regions 
originates primarily from anomalously close pairs of stripes.
This is demonstrated in Figs. \ref{doubledispfig}-\ref{doubleAfig}, which 
show that for an isolated bi-stripe, there is low-energy spectral weight 
in the vicinity of ($\pi/2,\pi/2$), even when the stripes are roughly quarter 
filled. The idea that the intensity of the nodal spectral weight is related 
to the degree of stripe disorder\cite{dual} conforms with the experimental 
finding (see Fig. \ref{arpes}) of more pronounced Fermi segments in the 
nodal direction in optimally doped LSCO, where the stripes do not statically 
order, than in LNSCO, where long-range stripe order is seen in neutron 
diffraction\cite{tranquada}.
Several other experimental observations can be understood in terms of 
our results, as discussed in Sec.~\ref{consequences}.

When residual interactions between electrons on a stripe are considered, 
each non-interacting localized band gives rise to a one-dimensional Luttinger 
liquid. As in the non-interacting case the two-dimensional $\vec k$ space 
structure, and especially the Fermi surface, remain prominent features of the
low-energy spectral response. 
However, along a cut in the BZ perpendicular to the Fermi surface, 
the spectral function mimics Luttinger liquid behavior. (See, e.g., 
the behavior along the nodal direction in Fig.~\ref{1dspectral}. ) 
This justifies the application of one-dimensional physics along 
directions that are not necessarily aligned with the stripes, as was 
done recently in Ref.~\ref{drorref}.

It is important to stress that we view the present results 
as reliable for exploring the qualitative effects of stripes on electronic 
structure, but not as a realistic study of the cuprate superconductors.
The parameters in our model have not been carefully adjusted to 
optimize any sort of fit to the data. We have certainly not included
any ``realistic'' band-structure effects, such as second neighbor hopping,
$t^{\prime}$, nor, except in Sec.~\ref{interactions}, considered any strong 
correlation effects, other than the stripes themselves. Phonons, dynamical 
stripe fluctuations, effects of transverse deformations of the stripe 
potential, dynamical magnetic fluctuations, and all other forms of of static 
or dynamical disorder are neglected in our calculations. They will all 
certainly have important consequences for the details of the measured 
electronic structures. However, it should also be stressed that, throughout 
this paper, we will be concerned with intermediate energies and length 
scales. Thus, for our purposes, the distinction between long-range and 
mesoscale stripe order, and between static and fluctuating stripes is 
{\em unimportant}, although of course these distinctions are essential 
at low energies and long wave-lengths.

\section{the model}
\label{model}

Since the high temperature superconductors are doped antiferromagnets, 
it is reasonable to expect slow fluctuations of a collective field 
representing the local staggered magnetization. One can think of this 
field as resulting from a Hubbard-Stratonovich transformation of the 
interacting many-body problem. However, we do not solve a Hartree-Fock 
theory for this field, i.e., we do not find the configuration that minimizes 
the Hartree-Fock energy. Instead we take as a minimal model a set of 
non-interacting electrons on a square lattice interacting with a static, 
staggered field which represents a characteristic ``snapshot'' of the field 
configuration:
\begin{eqnarray}
H=-&&\sum_{x,y,\sigma}(c^{\dagger}_{x,y,\sigma}c_{x+1,y,\sigma}+
c^{\dagger}_{x,y,\sigma}c_{x,y+1,\sigma}+{\mbox{H.C.}})\nonumber\\
+&&\sum_{x,y,\sigma}\sigma(-1)^{x+y}m(x,y)c^{\dagger}_{x,y,\sigma}
c_{x,y,\sigma} \; ,\label{H}
\end{eqnarray}
where $c_{x,y,\sigma}$ is the electron destruction operator at site 
$(x,y)$ and spin $\sigma=\pm$ and we have chosen units such that the hopping
matrix element and the lattice constant equal 1. Specifically, we consider 
the simplest possible ansatz to represent stripe configurations in which
\begin{equation}
m(x,y)=m\prod_{x_s}\Theta(x-x_s) \; ,
\label{wall}
\end{equation}
where $m$ is a constant, $\{x_s\}$ are a given set of positions of anti-phase 
domain walls, and where $\Theta$ is the antisymmetric step-function: 
$\Theta(x)=-\Theta(-x)=1$ for $x>0$ and $\Theta(0)=0$. This corresponds to 
an array of perfectly straight, site centered stripes of width one oriented 
in the $y$ direction. Depending on the choice of the set $\{x_s\}$ the 
potential can either be regular or disordered in the transverse ($x$) 
direction. 

Some features of the solution below do depend on details of this choice, 
such as whether the stripes are site or bond centered, whether they are of 
width 1 or wider, whether we include an additional collective field
that couples to the charge density \cite{salkola}, etc.; effects that we 
have explored, to some extent.  However, the important qualitative physics 
is apparent in this simplest of models, so we only report results for this 
model. Two more serious omissions, which are beyond the scope of the present 
paper, are the neglect of effects of the dynamical character of the 
collective fields, and the neglect of any shape deformations of the stripe 
order. The latter approximation implies that the electronic states are Bloch 
waves in the $y$ direction, with a wave-vector, $k_y$, which is 
conserved mod $\pi$ (due to the presence of the staggered field).  

Without the domain walls the system has an energy gap of magnitude 
$2m$ and two bands with energies $E=\pm\sqrt{m^2+{\epsilon}^2}$, with 
$\epsilon=-2[\cos(k_x)+\cos(k_y)]$. An isolated domain wall generates 
mid-gap states which are localized in the direction orthogonal to the wall,
as is shown analytically, below. 

For an array of domain walls, it would be exceedingly difficult to find 
analytic solutions for the mid-gap states. Instead we diagonalize such 
systems numerically for a given realization of the model. In each case, 
we take $m=1$ which corresponds to an ``intermediate coupling''
value of the ratio of the energy gap to the band-width of 2/8=1/4.
We study systems with finite width, $L_x$, in the $x$ direction where 
typically $L_x=320$; according to all tests we have applied, this is large 
enough to eliminate most finite size effects. Since $k_y$ is a good quantum 
number, the only place the finite size in the $y$ direction enters our 
calculations is when we perform sums over $k_y$; here, it is easy to show 
that the system sizes we have considered, $L_y=320$, are more than adequate 
to eliminate finite size effects.

\section{band--structure of a stripe array}

In this section, we consider the spectral weight distribution and band 
structure of a stripe array. If one ignores details concerning the matrix 
elements describing the photo-excitation process one can infer directly the 
single-particle spectral function, $A(\vec k,\omega)$, from ARPES.
For non-interacting electrons, $A(\vec k,\omega)$ can be expressed in 
terms of the exact single particle energy eigenstates 
$\Psi_{\alpha}(\vec r)$ and energies $E_{\alpha}$:
\ba
A(\vec k,\omega)&=& (L_{x}L_{y})^{-1}\sum_{\vec r,\vec r^{\prime}}
e^{i\vec k\cdot (\vec r-\vec r^{\prime})} \tilde A(\vec r,\vec 
r^{\prime},\omega) \; ,\nonumber \\
\tilde A(\vec r,\vec 
r^{\prime},\omega)&=& \sum_{\alpha}\Psi_{\alpha}(\vec r)\,
\Psi_{\alpha}(\vec r^{\prime}) \delta(\omega-E_{\alpha}) \; .
\ea   
     
\begin{figure}[ht]
\narrowtext
\begin{center}
\leavevmode
\noindent
\centerline{\epsfxsize=3.4in \epsffile{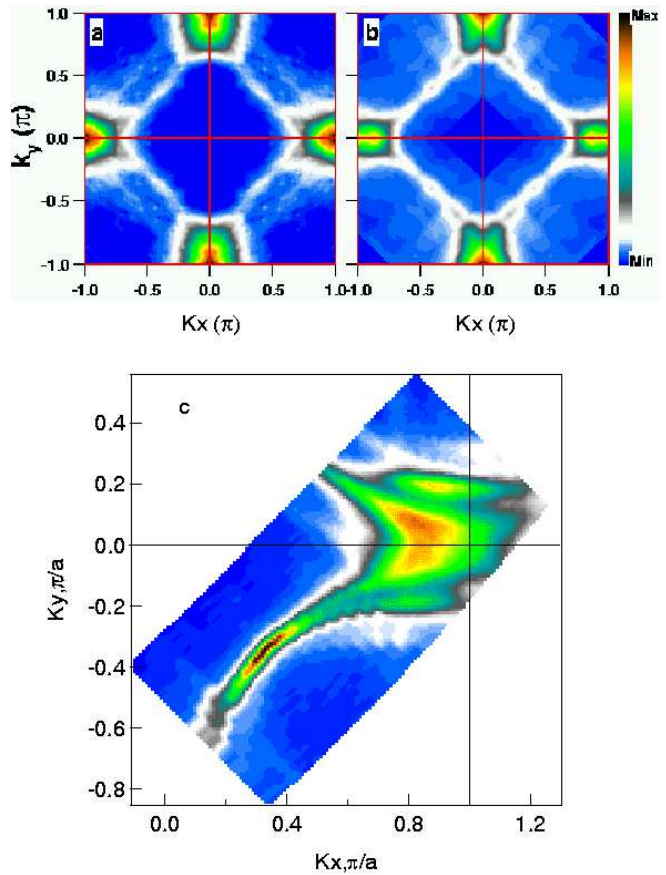}}
\end{center}
\caption{Distribution of low-energy spectral weight in the first Brillouin 
zone as measured by ARPES for various cuprates. The spectral weight was 
integrated over an energy window of 30meV below the Fermi energy. 
Results are shown for 
a) {\LNSCO} ($x=0.15$) measured at 15 K (from Ref. \protect{\ref{refdual}}), 
b) {\LSCO} ($x=0.15$) measured at 15 K (from Ref. \protect{\ref{refdual}}), 
and 
c) overdoped Pb-doped {\BSCCO} ($T_c$=70 K) measured at 20 K 
(from Ref. \protect{\ref{refpasha}}).}
\label{arpes}
\end{figure}

In order to characterize the momentum distribution of the low-energy 
spectral weight and reveal features such as the Fermi surface, it is 
appropriate to integrate the spectral function over a narrow energy 
window below the chemical potential $\mu$
\be
I(\vec k)=\int_{\mu-\Delta\omega}^{\mu} A(\vec k,\omega) d\omega \; .
\ee
The experimental data in Fig.~\ref{arpes} and the theoretical results 
in later figures are expressed in this way.

\begin{figure}[ht]
\narrowtext
\begin{center}
\leavevmode
\noindent
\centerline{\epsfxsize=3.2in \epsffile{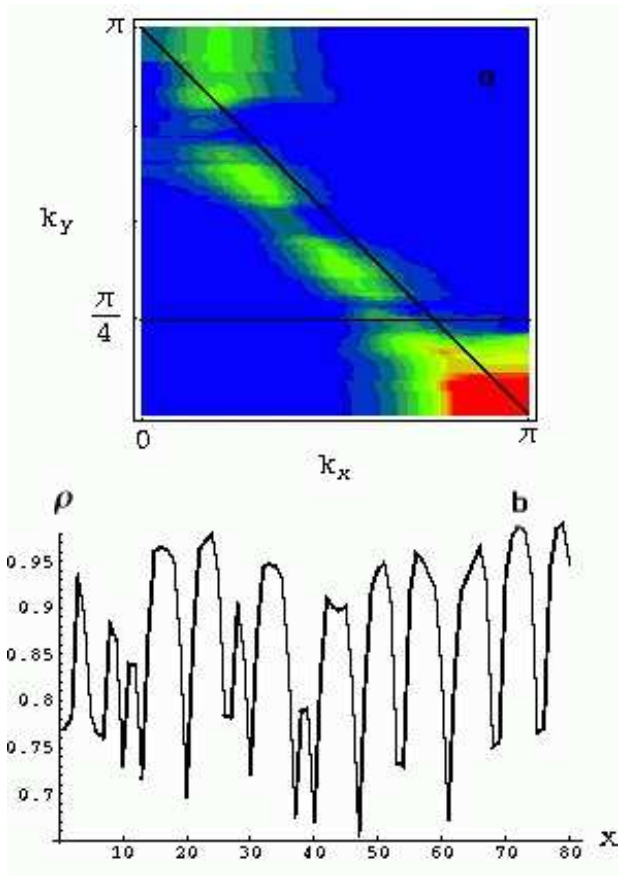}}
\end{center}
\caption{ a) The low-energy spectral weight in the first quadrant of the 
Brillouin zone integrated over an energy interval $\Delta\omega=0.2$ below 
$\mu$ for a disordered stripe array with mean spacing $\ell=4$ and $m=1$ 
at $\delta=1/8$ doping. The system size is $320\times 320$ and the amplitude 
color is coded from blue through green to red, 
as in Fig. \protect{\ref{arpes}}. 
b) Typical electron density profile of the array as a function of $x$ 
(perpendicular to the stripe axis). While some of the dips correspond to 
stripe positions, others, such as the one at $x\approx53$, correspond to 
bi-stripes. Thus, in comparison to Fig. \protect{\ref{fig418}}, below, the 
density of dips is smaller although the density of stripes is the same.}
\protect{}
\label{random}
\end{figure}

\begin{figure}[h]
\narrowtext
\begin{center}
\leavevmode
\noindent
\centerline{\epsfxsize=3.1in \epsffile{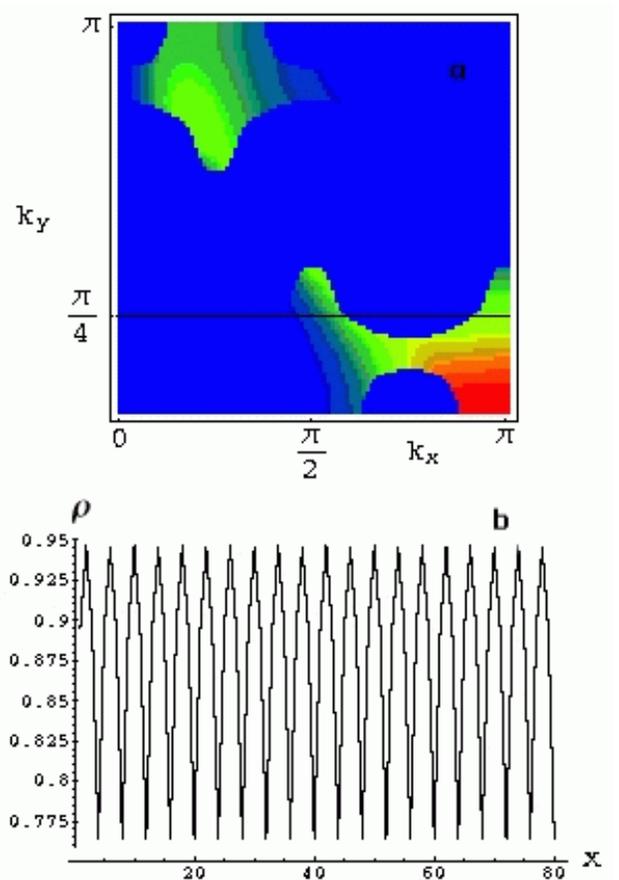}}
\end{center}
\caption{Same as in Fig. \protect{\ref{random}} but for an ordered stripe 
array with $\ell=4$, $\delta=1/8$ and $\Delta\omega=0.2$.}
\label{fig418}
\end{figure}

In Fig.~\ref{fig418} we report results for an ordered array with a spacing 
between stripes of $\ell=4$. In Fig.~\ref{random}, we do the same for an 
array with a distribution of stripe spacings chosen randomly between 1 and 7, 
i.e., a flat distribution with mean $\ell=4$. In both cases, we have fixed 
$\mu$ such that the average electron density per site is $1-\delta$ 
where $\delta=1/2\ell$. ($\delta$ is known as the density of ``doped holes.'')
This corresponds {\it on the average} to quarter filled stripes, such that 
for $\ell=4$ we obtain $\delta=1/8$. In these figures, we have chosen 
$\Delta\omega=0.2$, but the qualitative character of the distribution is not 
highly sensitive to this choice. The data for the disordered array is for 
a given realization of the stripe distribution, but the system is large 
enough that the results are self averaging in the sense that the 
corresponding figures look similar to the eye for different realizations.

Clearly, a number of salient features of the low-energy spectral weight 
of these various stripe arrays are reminiscent of those measured in
ARPES. In particular the disordered array, Fig.~\ref{random}, looks 
strikingly like the corresponding spectrum in LSCO. In both cases,
most of the spectral weight is in a broad, flat region close to the 
$(\pi,0)$ point with a fainter image near $(0,\pi)$. For the ordered 
array this exhausts the low-energy spectral weight, but for the disordered 
array, as in the experiments, a small amount of spectral weight lies 
along what would have been the band-structure Fermi surface in the 
nodal region. (Of course, in order to compare the results more closely 
with the ARPES measurements we should symmetrize these results around 
$\kp=\ko$ to allow for stripes running in both directions in different 
domains.) 

These results demonstrate that two-dimensional structures, including 
an apparent Fermi surface, can appear in the ARPES data of a system which is 
dynamically one-dimensional. It is also noteworthy that $y$ directed stripes 
not only lead to substantial low energy spectral weight in the anti-nodal 
region near $(0,\pi)$, as naively expected, but also in the reciprocal 
region, near $(\pi,0)$, as well. (This observation is important in systems 
with macroscopic stripe orientational order, including, potentially, 
YBCO \cite{anisotropy}.) We also note that for the ordered array, there are 
oscillations in the spectral weight as a function of $k_x$. It is apparent 
from Fig.~\ref{fig418} that these oscillations reflect the periodicity of 
the array ($2\pi/4=\pi/2$ in this case. We have checked \cite{web} that for 
longer period arrays, for example with $\ell=6$, the period of these 
oscillations shifts accordingly.) That such oscillations have been observed 
in LNSCO \cite{lnsco} further corroborates the stripe interpretation 
of the results. 

\begin{figure}[h]
\narrowtext
\begin{center}
\leavevmode
\noindent
\centerline{\epsfxsize=3.2in \epsffile{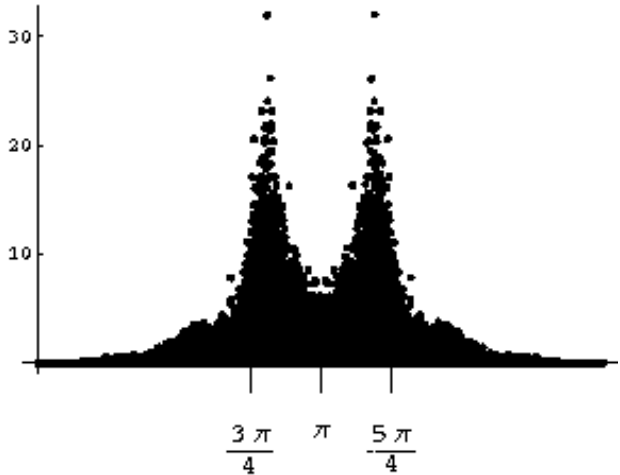}}
\end{center}
\caption{The squared magnitude of the Fourier transform of the field 
$m(x,y)$ for a disordered array of size $L_x=100,000$. This quantity is 
proportional to the spin-spin correlation function measured by neutron 
scattering.}
\protect{}
\label{diffraction}
\end{figure}

We have also calculated the electronic density distributions of the 
disordered and ordered arrays and they are shown next to the spectral maps 
in Figs.~\ref{random} and \ref{fig418}. We see that in the ordered 
case, the density modulations have the same periodicity as the 
stripe potential, but in the disordered array, the mean spacing between 
troughs (or peaks) of the density is larger than the mean spacing between 
stripes. This reflects the fact that only a single density depression 
occurs where two stripes are close together. To demonstrate this fact 
and make connection with neutron diffraction experiments, we have computed 
structure factors from the Fourier transforms of the densities of electronic 
charge ($\rho$) and z-spin component of the spin ($S_z$). 
For the ordered array of period 4 the charge signal is strongly peaked 
at $k_x=0$ with a small satellite at $k_x=\pi/2$ and the spin signal is 
peaked at $k_x=3\pi/4$ and $k_x=5\pi/4$. For the disordered array the charge 
and spin peaks are shifted towards $k_x=0$ and $k_x=\pi$ respectively. 
In Fig.~\ref{diffraction} we show the Fourier transform of the 
autocorrelation function of the random field $m$. (We show the results 
for $m$ rather than for $S_z$, because it can be computed for very large 
arrays, where the result is self-averaging. However, we have verified that 
the two quantities give similar results\cite{web}.) 
The fact that for the disordered array, the spin peaks are not only 
broadened, but are shifted towards the ($\pi,\pi$) point seems to be 
generic behavior for disordered arrays.

\begin{figure}[h]
\narrowtext
\begin{center}
\leavevmode
\noindent
\centerline{\epsfxsize=3.32in \epsffile{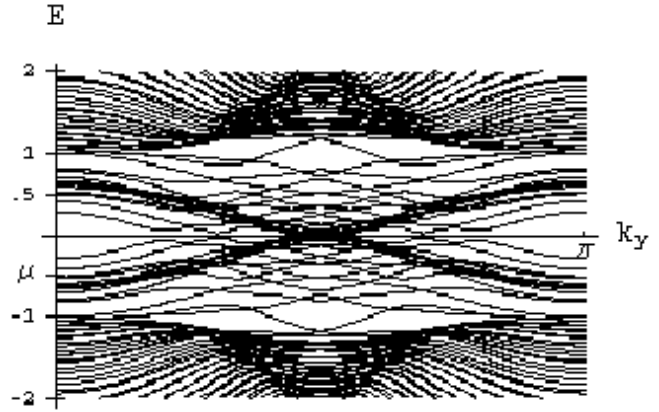}}
\end{center}
\caption{Band-structure of a disordered array with mean stripe spacing 
$\ell=4$ and $m=1$. $\mu$ is the $T=0$ chemical potential for a density 
of doped holes $\delta=1/8$. The system size is  80$\times$320}
\label{randomdispfig}
\end{figure}

Since $k_{y}$ is still a good quantum number, we can still talk 
about a (one-dimensional) band-structure of the disordered array, 
as shown in Fig.~\ref{randomdispfig}. However, as there are 
precisely $2L_{x}$ bands, this figure becomes rather dense for a 
large number of stripes, so we have purposely reduced the system size 
to $L_{x}=80$ and the energy window to $|E|<2$ for graphical clarity. 
We devote the next section to a closer examination of this band-structure. 

Because for each value of $k_{y}$, the effective Hamiltonian is that 
of a one-dimensional disordered system in the $x$ direction, we know a 
priori that all the states are exponentially localized transverse to 
the stripes. However, for the states that lie within the Mott gap, 
$|E|<m=1$, one might expect the states to be highly localized in 
the neighborhood of one or two stripes, while for $|E|>m=1$, the 
states are more likely extended scattering states, that are only 
weakly localized. That this expectation is realized can be seen
in at least three ways. It is clear from Fig.~\ref{randomdispfig}, 
that there is a dense set of states at $E>m=1$, whereas for $E<m=1$ 
there is a spider web of identifiable one-dimensional bands; the 
discreteness of these bands is a reflection of their large degree of 
localization in the transverse direction. By plotting the states
in real space along $x$ we have confirmed that all the states in the gap
decay exponentially in a few lattice constants, whereas the scattering
states are essentially delocalized. Finally, 
we have computed the participation ratio, $P_{\alpha}\equiv \sum_{\vec 
r}|\Psi_{\alpha}(\vec r)|^{4}\approx (2\xi)^{-1}$, where $\xi$ is the 
localization length. For a system of size $320\times 320$, we find $P$s 
for the states with $|E|>m$ of order $P\sim 1/100$ (which might get even 
smaller for larger system sizes), while for $|E|<m$, typically $P\approx 
0.1-0.25$. For moderate doping $\delta$ the Fermi energy always lies 
inside the set of highly localized bands. (See for example, 
Fig.~\ref{randomdispfig} where the value of the Fermi energy corresponding 
to $\delta=1/8$ is indicated.) It is clear that all of the spectral weight 
shown in Fig.~\ref{random} comes from these localized, and hence dynamically 
entirely one-dimensional states.

\section{Isolated stripes and bi-stripes}

In order to get a better understanding of the nature of the localized 
states and how they give rise to the very characteristic two-dimensional 
distribution of spectral weight, we look at some simpler problems consisting 
of a single or a pair of proximate domain walls. 

Consider the situation in which the defects, i.e., one or more stripes, are
confined to a region about $x=0$. It is straightforward to explicitly write 
the wavefunctions for fixed $k_y$ in the asymptotic regions to the left and 
right of the defect; the eigenvalue problem is then solved by matching these 
solutions across the defect region. For $|E|<m$, the asymptotic states are 
exponentially falling functions of $x$. For instance, to the right of the 
defect, the Hamiltonian (\ref{H}) is diagonalized by
\begin{eqnarray}
\phi_{k_y}(x,y)=
e^{-\kappa x}e^{i(qx+k_yy)}\left[ 1+e^{i\alpha}e^{i\pi(x+y)}\right],
\label{singlewave}
\end{eqnarray}
where the quantities $\kappa$, $q$, and $\alpha$ are the implicit functions 
of $k_y$ and the energy $E$ which (for $\sigma=1$) satisfy
\begin{eqnarray}
0&=&\cos (q)\cosh(\kappa)+\cos (k_y) \; ,\label{constraint} \\
E^2&=&m^2-4\sinh^2(\kappa)\sin^2(q) \; ,\label{energy} \\
e^{i\alpha}&=&[E+2i\sinh(\kappa)\sin(q)]/m \; . \label{alpha}
\end{eqnarray}
It follows from Eqs. \ref{constraint}-\ref{alpha} that if (\ref{singlewave}) 
is a solution then so is (\ref{singlewave}) with $q\rightarrow -q$,  
$\alpha\rightarrow -\alpha$ and the same energy. Thus the total 
wavefunction $\Phi_{k_y}(x,y)$ is a linear superposition of the two. 
To the left of the defect the same equations hold after substituting 
$\kappa\rightarrow -\kappa$ and, in the case of an anti-phase defect, 
$m\rightarrow -m$. A fourth equation, which depends on the nature of 
the defect, is obtained by integrating the Schroedinger equation across 
the defect region and defines the eigenvalue problem, $E=E(k_y)$. 
(If the defect region is sufficiently broad, there may be multiple 
solutions to the eigenvalue equation, corresponding to multiple 
mid-gap bands.)

Much of the interesting physics is already implicit in these relations. The
distribution of spectral weight in the two dimensional BZ associated with a 
state of given crystal momentum $k_y$ is given by $|\Psi_{k_y}(k_x)|^2$, 
where $\Psi$ is the Fourier transform of $\Phi$. Clearly, this weight is 
peaked near $k_x=q(k_y)$ in a region of width $\Delta k_x\approx\kappa(k_y)$. 
For small $\kappa$, Eq. (\ref{constraint}) implies that $q(k_y)$ simply 
traces out the underlying ``diamond'' Fermi surface given by the unperturbed 
band-structure.

\begin{figure}[h]
\narrowtext
\begin{center}
\leavevmode
\noindent
\centerline{\epsfxsize=3.2in \epsffile{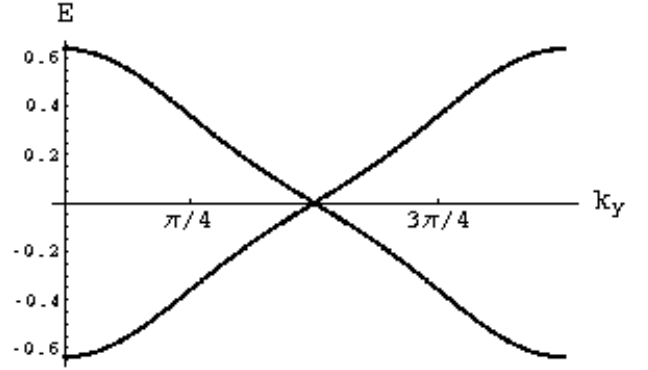}}
\end{center}
\caption{Dispersion $E(\kp)$ of the bound states of a single anti-phase
domain wall with $m=1$.}
\label{singledispfig}
\end{figure}

\begin{figure}[h]
\narrowtext
\begin{center}
\leavevmode
\noindent
\centerline{\epsfxsize=3.2in \epsffile{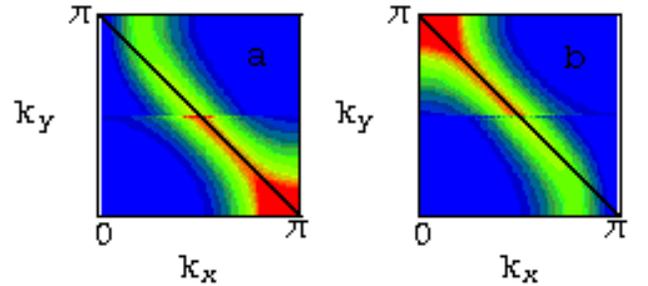}}
\end{center}
\caption{Integrated spectral weight over all negative (a) and positive (b) 
energy bound states of a single anti-phase domain wall with $m=1$. 
The apparent discontinuity across $k_{y}=\pi/2$ is due to 
the band crossing.}
\label{singleAfig}
\end{figure}

To be explicit, let us consider the case of a single site-centered 
anti-phase domain wall, ($\downarrow \uparrow \downarrow \uparrow \cdot 
\downarrow \uparrow\downarrow\uparrow$) of the type which appears in Eq.  
(\ref{wall}). In this case, the eigenvalue equation is readily derived
\be
E(k_y)=\pm2\tanh^2(\kappa)\cos(\kp).
\ee
The resulting band-structure and spectral weight distribution in the BZ is 
shown in Figs. \ref{singledispfig} and ~\ref{singleAfig}. Note that the 
mid-gap band has energy zero precisely where $k_y=\pi/2$, so that for a 
hole-doped stripe, for which the Fermi energy will certainly lie at negative 
energies, there will never be low-energy spectral weight at the nodal point. 
A generic feature of the spectral weight distribution that is apparent in 
the figure is the fact that the negative energy states have more weight in 
the lower-half of the first quadrant of the BZ than in the upper-half. 
This feature is seen very clearly also in the disordered array problem 
(Fig.~\ref{random}). 

It is particularly illuminating to examine this result in the small $m$ 
limit. The four implicit equations can be solved to leading order in $m$ 
to obtain (for $0< k_y < \pi$)
\ba
q(k_y) &=& \pi-k_y \; , \\
\kappa(k_y)&=&\frac {m}{2\sin(k_y)} \; , \\
E(k_y)&=&\pm 2\kappa^{2}(k_y) \cos(k_y) \; .
\ea 
As advertised, $q(k_y)$ precisely traces out the diamond Fermi surface. 
$\kappa^{-1}(k_y)\sim v_F(k_y)/m$ is the extent of the wavefunction in 
real space, implying that $|\Psi_{k_y}(k_x)|^2$ is quite sharply peaked 
near $k_x=q(k_y)$ in the nodal regions, where the Fermi velocity, $v_F$, 
is large, and is much more diffused in the anti-nodal regions, where 
$v_F\rightarrow 0$.  (Where $v_{F}=0$, the expression for $\kappa$ breaks 
down and we find $\kappa\sim\sqrt{m}$.) These features are, to a large extent, 
independent of the nature of the defect, so long as $m$ is not too large. 
It is also worth noting that the energy width of the mid-gap band is of order 
$m^2$, although the gap itself is 2$m$, and hence much larger.

The single anti-phase domain wall solution 
captures most of the general features of the array of domain walls:  
The spectral weight is concentrated about the diamond Fermi surface, 
it is largest in a broad anti-nodal region, and is larger in the 
lower-half than in the upper-half of the BZ quadrant. Indeed, the
low-energy spectral distribution of the ordered stripe arrays is, to first 
approximation, simply the sum of contributions of isolated stripes, 
even when the stripe spacing is only four lattice constants.

However, the disordered stripe array is not simply a superposition of single 
wall states, especially in the nodal region. In fact, while it is clear 
from the band-structure of the disordered array shown in 
Fig.~\ref{randomdispfig}, that a large fraction of the mid-gap bands are  
very similar to the mid-gap band of a single isolated stripe  
(Fig.~\ref{singledispfig}), there are other bands, especially those that 
dominate the spectrum at $k_{y}=\pi/2$ and $E$ near $\mu$, that look
very different. 

\begin{figure}[th]
\narrowtext
\begin{center}
\leavevmode
\noindent
\centerline{\epsfxsize=3.2in \epsffile{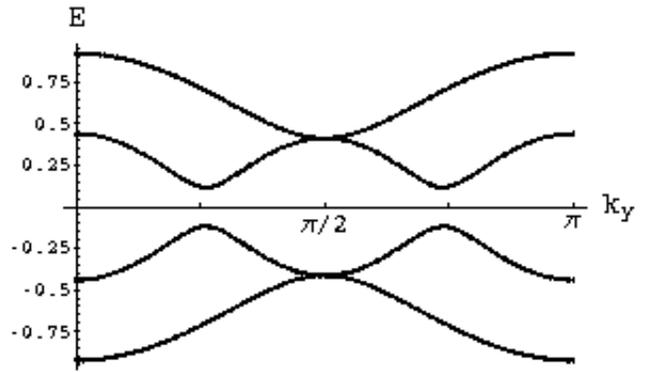}}
\end{center}
\caption{Spectrum $E(\kp)$ of the bound states of an effective
in-phase domain wall consisting of two one site anti-phase domain walls one 
site apart ($\uparrow \downarrow \uparrow \downarrow \cdot \uparrow \cdot 
\downarrow\uparrow \uparrow \downarrow$) with $m=1$.}
\label{doubledispfig2}
\end{figure}

\begin{figure}[h]
\narrowtext
\begin{center}
\leavevmode
\noindent
\centerline{\epsfxsize=3.2in \epsffile{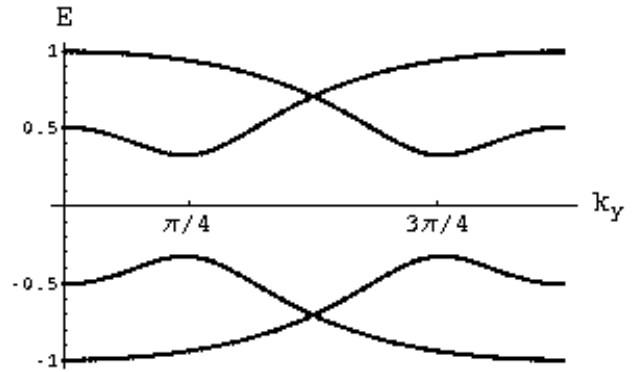}}
\end{center}
\caption{Spectrum $E(\kp)$ of the bound states of a two site wide in-phase 
domain wall ($\uparrow \downarrow \uparrow \downarrow \cdot \ \cdot \uparrow
\downarrow \uparrow\downarrow$) with $m=1$.}
\label{doubledispfig}
\end{figure}

Guided by the observation that the nodal weight is missing for an ordered 
array, it is reasonable to assume that this feature is related to local 
configurations with two or more domain walls in close proximity. 
Consider therefore the problem of two anti-phase domain walls in close 
proximity. Clearly, the statement of proximity is related to the decay length 
$\xi =1/\kappa$ of the single domain wall problem. If the distance between 
the walls is greater than $\xi$, the walls can be considered independent, 
while for smaller distances they interfere. For $m=1$ we find 
$1/\kappa\approx 2$ and only weakly dependent on $\kp$, implying that 
domain walls separated by more then two sites are roughly independent. 
Fig.~\ref{doubledispfig2} shows the energy spectrum for the bound states 
of two anti-phase domain walls separated by one site. The spectrum is gapped 
and it is clear that doping such a system can give rise to low-energy 
spectral weight around $\kp=\pi/2$. In Fig.~\ref{doubledispfig} we repeat 
the same calculation for two proximate stripes. 

\begin{figure}[h]
\narrowtext
\begin{center}
\leavevmode
\noindent
\centerline{\epsfxsize=3.2in \epsffile{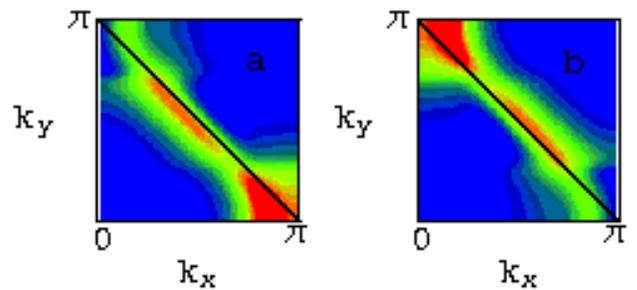}}
\end{center}
\caption{Integrated spectral weight over all negative (a) and positive (b) 
energy states of an in-phase domain wall consisting of two one site 
anti-phase domain walls one site apart with $m=1$.}
\label{doubleAfig}
\end{figure}

Particularly interesting is the distribution of spectral weight in momentum 
space for this system as shown in Figure~\ref{doubleAfig}. As with the single 
wall problem, the weight hovers around the band-structure Fermi surface, but 
with the distinct feature that the weight of the negative energy states is 
shifted towards the zone center whereas the positive energy states are 
shifted towards the $(\pi,\pi)$ point. We find that the magnitude of this 
shift increases with $m$. Intuitively it is of course not surprising that 
the negative energy states are shifted towards the zone center where the 
tight-binding energies are lower and conversely for the positive energy 
states. 


\section{including interactions}
\label{interactions}

So far we have been concerned with the distribution of spectral weight 
in the non-interacting model defined in Eq. (\ref{H}) when holes are 
introduced into its single-particle levels. Including the effects of 
interactions between electrons in a doped stripe is a natural extension 
of this model. In the following we consider this issue for the case of 
the disordered stripe array. 

To the extent that the localized bands shown in Fig. \ref{randomdispfig} 
can be considered as independent we may calculate the low-energy and 
long-wavelength spectral response of the system as a sum over interacting 
Luttinger liquid contributions $A_{1d}(k,\omega)$
\be
A(\vec k,\omega)=\sum_{k_F}|\Psi_{k_F}(k_x)|^2
A_{1d}[k_y-k_F,\omega;\gamma(k_F)] \; .
\label{totalA}
\ee
Here the sum runs over the localized bands that cross the Fermi energy.
Every crossing point is a Fermi point $k_F\equiv \vec k_F\cdot \hat y$
for the one-dimensional channel defined by the band. Each channel is 
also characterized by the $k_x$ profile of its wavefunction 
$\Psi_{k_F}(k_x)$ and its Luttinger parameters $K_{c,s}$ and velocities
$v_{c,s}$ which are collectively denoted as $\gamma(k_F)$ in 
(\ref{totalA}). 

The evaluation of the spectral function (\ref{totalA}) simplifies in the 
small $m$ limit where the spectral weight is concentrated (within a width 
of order $m$) along the diagonal connecting the anti-nodal points and is 
roughly constant. The total spectral function is then proportional to 
$A_{1d}(k_x+k_y-\pi,\omega;\gamma)$, where we have assumed that $\gamma$ 
is the same for all bands. As we see, in this case, the spectral function 
exhibits one-dimensional behavior over the entire BZ. This description is 
valid as long as the stripe array is sufficiently dilute such that the 
overlaps between different localized bands are small. 

\begin{figure}[th]
\narrowtext
\begin{center}
\leavevmode
\noindent
\centerline{\epsfxsize=3.2in \epsffile{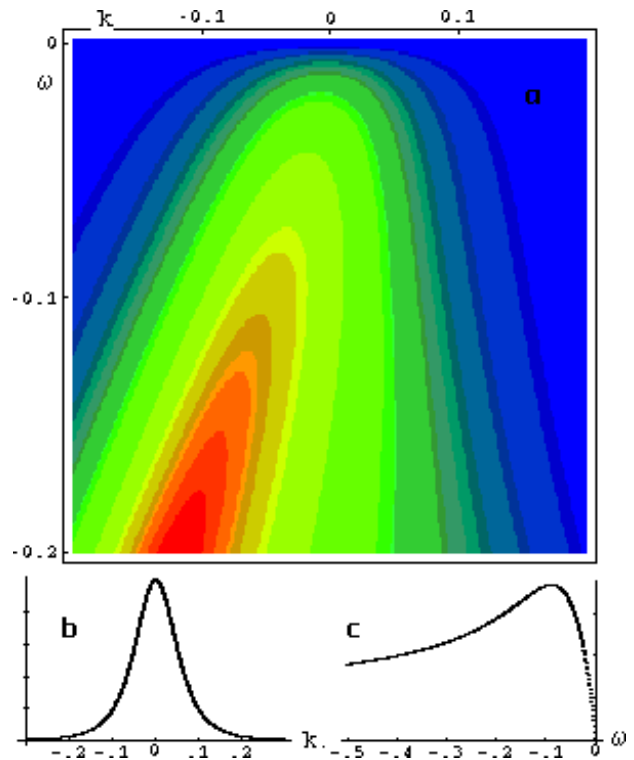}}
\end{center}
\caption{a) The zero-temperate spectral function $A(k,\omega)$ calculated 
from Eq. (\ref{totalA}) assuming a Lorentzian distribution of spectral 
weight centered on a line parallel to the anti-diagonal. Here $v_c=v_s=1$, 
$K_s=1$, $\gamma_c=(K_c+K_c^{-1}-2)/8=0.3$ and $\kappa=0.5$ (which is what 
we find for a single stripe with $m=1$.) $k$ is measured along the BZ
diagonal and relative to the Fermi point near $(\pi/2,\pi/2)$, $\omega$ 
is measured relative to the chemical potential $\mu$. b) MDC at $\omega=0$, 
c) EDC at $k=0$.}
\label{1dspectral}
\end{figure}

For intermediate values of $m$ performing the convolution in (\ref{totalA}) 
is difficult owing to the complicated distribution of the low-energy spectral 
weight in momentum space. However, the spectral function continues to mimic 
one-dimensional behavior along directions that are not necessarily aligned 
with the stripes. To demonstrate this point we consider the spectral function 
along a cut in the BZ that runs in the nodal direction [from $(0,0)$ to 
$(\pi,\pi)$]. Since the zero-temperature $A_{1d}(k-k_F,\omega;\gamma)$ 
vanishes unless $|\omega|>{\rm min}(v_c,v_s)|k-k_F|$ \cite{dror} it is 
clear that the important contribution to $A(\vec k,\omega)$ along this 
cut, and at low energies, comes from bands that cross the Fermi energy 
near $k_y=\pi/2$. Moreover, we have shown that these bands produce  
spectral weight that is concentrated along a line parallel to the 
anti-diagonal [$(\pi,0)$ to $(0,\pi)$]. Therefore, for our purpose, 
we may approximate the spectral distribution by a Lorentzian distribution 
centered on this line with half-width at half-maximum of $\kappa$. 
In Fig.~\ref{1dspectral} we present the spectral function which results 
from Eq. (\ref{totalA}) under the assumption of such weight distribution. 

The spectral function in Fig.~\ref{1dspectral} shows a close resemblance 
to one-dimensional spectral functions (compare for example with Ref. 
\ref{drorref}). In particular it exhibits a striking dichotomy between 
sharp MDCs (momentum distribution curves - cuts at constant $\omega$) 
and broad EDCs (energy distribution curves - cuts at constant $k$), 
which we believe is a telltale of electron fractionalization \cite{dror}.   
It is interesting to note, however, that $A(\vec k,\omega)$ in 
Fig.~\ref{1dspectral} is smoother than the corresponding zero-temperature 
one-dimensional spectral function and does not diverge at $\omega=vk$ for 
$k<0$ like the latter. This divergency is smeared by the convolution in Eq. 
(\ref{totalA}) and the result is a spectral function that resembles 
$A_{1d}(k,\omega)$ but at a finite temperature. 

The present analysis is the first which combines the remarkable non-Fermi 
liquid features of the one-dimensional electron gas with the two-dimensional 
structure of a Fermi surface. At a qualitative level, it justifies the 
analysis presented in Ref. \ref{drorref}, in which the measured spectral 
functions of the high temperature superconductors was analyzed in terms of 
the corresponding expressions for a one-dimensional Luttinger liquid. 
In particular, we have presented here a microscopic rationale, which was 
missing at the time of that earlier work, for the existence of approximate 
one-dimensional dynamics in the nodal region of the Fermi surface.

\section{Further Consequences}
\label{consequences}

In addition to the gross similarities between the experimentally 
observed spectral functions in LSCO, LNSCO and BSCCO and those obtained from 
simple stripe models, there are a number of other aspects of the present 
results that are potentially relevant to the interpretation of experiment.

A phenomenon that is fairly generally observed in ARPES measurements on high 
temperature superconductors is that a sharp quasi-particle like peak appears 
in the spectrum in the anti-nodal regions as the temperature is 
reduced below $T_c$. Recent experiments on untwinned \YBCO (YBCO) crystals 
have found that the spectral-weight of this sharp feature
exhibits strong anisotropy between the different anti-nodal regions 
of the BZ \cite{anisotropy}. There are good empirical \cite{mook2,ando} 
and theoretical\cite{nature} reasons to believe that the stripes in this 
material are partially oriented by the chain potential on the Cu-O planes. 
Thus, a large anisotropy in plane-related features is to be expected. 
Moreover, we have previously shown\cite{erica} that the emergence 
of such a peak is naturally explained as a dimensional crossover in
a quasi-one dimensional superconductor. However, the presence of low-energy 
spectral weight in the anti-nodal region transverse to the stripes seems 
surprising at first. Nevertheless, as noted above, stripes induce low-energy 
spectral weight in both anti-nodal regions, although with different spectral 
weights - this is a consequence of strong mixing between states separated by 
the antiferromagnetic wave-vector, $(\pi,\pi)$. The observed anisotropy is 
reproduced naturally by the present model - see Fig.~\ref{random}.

One would generally expect electrons diffracting off an ordered stripe array 
to exhibit intensity modulations with the corresponding period, as 
indeed is apparent in the period $\pi/2$ structure seen in Fig.~\ref{fig418}.
We believe, in agreement with the interpretation proposed Zhou et al. 
\cite{lnsco}, that this effect has already been seen in ARPES 
spectra of LNSCO \cite{lnsco}. 
Since all previous observations of charge-density wave ordering are indirect 
(e.g. they detect the accompanying lattice distortion density wave), this 
observation is potentially of more far reaching importance as a method for 
detecting charge order.

Experiments on both electron and hole-doped cuprates indicate that 
the chemical potential remains within the Mott gap, and is a 
remarkably slowly varying function of dopant concentration; instead,
new dopant induced states appear in the gap\cite{allen,watanabe,ino}. 
Of course, this behavior is characteristic of the sort of stripe models 
envisaged here, in which the density of stripes increases with dopant 
concentration so as to accommodate the added charges in mid-gap states.
This is a form of ``topological doping'' \cite{topological} analogous to 
soliton doping of polyacetylene \cite{chx}. As is the case in polyacetylene, 
we expect a variety of additional aspects of topological doping to be 
manifest in experiments on the cuprates, including dopant induced 
mid-gap optical absorption\cite{salkola,machida}, 
dopant induced infra-red active phonon modes\cite{basov}, etc.


Finally, the observation that the structure factor of a disordered 
stripe array exhibits incommensurate peaks that are not only broadened 
by disorder, but also shifted to smaller incommensurability 
(Fig.~\ref{diffraction}) may have other interesting implications. 
Inelastic neutron scattering experiments in LNSCO\cite{tranquada}, 
LSCO \cite{yamada} and YBCO \cite{mook} reveal incommensurate
magnetic peaks displaced from the antiferromagnetic wave vector, 
$(\pi,\pi)$, by an amount which first grows roughly linearly with
doping, but then tends to saturate (typically beyond $\delta=1/8$).
This is generally taken as evidence that the concentration stripes, 
likewise, first increases linearly and then saturates. Our present 
results suggest that, at least in part, the saturation of the 
incommensurability may reflect increasingly strong stripe fluctuations, 
rather than a sudden saturation of the stripe concentration.

\acknowledgements
We would like to acknowledge helpful discussions with J.~W.~Allen, 
P.~D.~Johnson, J.~M.~Tranquada and Z.-X.~Shen. We thank P.~Bogdanov 
and X.~J.~Zhou for providing us with their data. M.~Granath acknowledges 
support from the Swedish Foundation for International Cooperation in 
Research and Higher Education (STINT). S.~A.~Kivelson, V.~Oganesyan 
and D.~Orgad were supported in part by NSF grant No. DMR-0110329 and 
DOE grant No. DE-FG03-00ER45798.

\end{multicols}

\begin{references}

\bibitem{salkola}  M.~Salkola, V.~J.~Emery and S.~A.~Kivelson, 
\prl\ {\bf 77}, 155 (1996). \label{salkolaref}

\bibitem{hanke} M.~G.~Zacher, R.~Eder, E.~Arrigoni and W.~Hanke, 
Phys. Rev. Lett. {\bf 85}, 2585 (2000).

\bibitem{machida} M.~Ichioka and K.~Machida, J. Phys. Soc. Jpn. {\bf 68}, 
4020 (1999).

\bibitem{lnsco} X.~J.~Zhou, P.~Bogdanov, S.~A.~Kellar, T.~Noda, H.~Eisaki, 
S.~Uchida and Z.-X.~Shen, Science {\bf 286}, 268 (1999).

\bibitem{erica} E.~W.~Carlson, D.~Orgad, S.~A.~Kivelson and V.~J.~Emery, 
Phys. Rev. B {\bf 62}, 3422 (2000).

\bibitem{dror} D.~Orgad, S.~A.~Kivelson, E.~W.~Carlson, V.~J.~Emery, 
X.~J.~Zhou and Z.-X.~Shen, Phys. Rev. Lett {\bf 86}, 4362 (2001). 
\label{drorref}

\bibitem{dual} X.~J.~Zhou, T.~Yoshida, S.~A.~Kellar, P.~V.~Bogdanov,
E.~D.~Lu, A.~Lanzara, M.~Nakamura, T.~Noda, T.~Kakeshita, H.~Eisaki, 
S.~Uchida, A.~Fujimori, Z.~Hussain and Z.-X.~Shen, Phys. Rev. Lett. 
{\bf 86}, 5578 (2001). \label{refdual}

\bibitem{realistic} M.~Fleck, E.~Pavarini and O.~K.~ Andersen, 
cond-mat/0102041 (unpublished).

\bibitem{valla} T.~Valla, A.~V.~Fedorov, P.~D.~Johnson, Q.~Li, G.~D.~Gu 
and N.~Koshizuka, Phys. Rev. Lett. {\bf 85}, 828 (2000). 

\bibitem{tranquada} J.~M.~Tranquada, J.~D.~Axe, N.~Ichikawa, 
A.~R.~Moodenbaugh, Y.~Nakamura and S.~Uchida, Phys. Rev. Lett. {\bf 78}, 
338 (1997). 

\bibitem{pasha} P.~V.~Bogdanov, A.~Lanzara, X.~J.~Zhou, S.~A.~Kellar, 
D.~L.~Feng, E.~D.~Lu, H.~Eisaki, J.-I.~Shimoyama, K.~Kishio, Z.~Hussain 
and Z.-X.~Shen, cond-mat/0005394 (unpublished). \label{refpasha}

\bibitem{anisotropy} D.~H.~Lu, D.~L.~Feng, N.~P.~Armitage, K.~M.~Shen, 
A.~Damascelli, C.~Kim, F.~Ronning, Z.-X.~Shen, D.~A.~Bonn, R.~Liang, 
W.~N.~Hardy, A.~I.~Rykov and S.~Tajima, Phys. Rev. Lett. {\bf 86}, 
4370 (2001). 

\bibitem{web} Data files for the figures presented in this paper, as well
as additional results mentioned in the text, but not presented in the 
paper, can be found at http://fy.chalmers.se/$\sim$granath/figures/. 
In particular, the low-energy spectral distribution for an ordered array 
of period $\ell=6$, and the structure factors for $S_{z}$ and $\rho$ for the 
disordered array with mean stripe spacing $\ell=4$ are available, 
there, for viewing.

\bibitem{mook2} H.~A.~Mook, P.~Dai, F.~Do\u{g}an, and R.~D.~Hunt, 
Nature (London) {\bf 404}, 729 (2000).

\bibitem{ando}  Y.~Ando, K.~Segawa, S.~Komiya and A.~N.~Lavrov, 
cond-mat/0108053 (unpublished).

\bibitem{nature} S.~A.~Kivelson, E.~Fradkin and V.~J.~Emery, Nature (London) 
{\bf 393}, 550 (1998).
 
\bibitem{allen} J.~W.~Allen, C.~G.~Olson, M.~B.~Maple, J.-S.~Kang, L.~Z.~Liu, 
J.-H.~Park, R.~O.~Anderson, W.~P.~Ellis, J.~T.~Market, Y.~Dalichaouch and 
R.~Liu, Phys. Rev. Lett. {\bf 64}, 595 (1990); R.~O.~Anderson, R.~Cleassen, 
J.~W.~Allen, C.~G.~Olson, C.~Janowitz, L.~Z.~Liu, J.-H.~Park, M.~B.~Maple, 
Y.~Dalichaouch, M.~C.~de~Andrade, R.~F.~Jardim, E.~A.~Early, S.-J.~Ho and 
W.~P.~Ellis, Phys. Rev. Lett. {\bf 70}, 3163 (1993).

\bibitem{watanabe} T.~Watanabe, T.~Takahashi, S.~Suzuki, S.~Sato and 
H.~Katayama-Yoshida, Phys. Rev. B {\bf 44}, 5316 (1991). 

\bibitem{ino} A.~Ino, C.~Kim, M.~Nakamura, T.~Yoshida, T.~Mizokawa, 
Z.-X.~Shen, A.~Fujimori, T.~Kakeshita, H.~Eisaki and S.~Uchida, Phys. Rev. B 
{\bf 62}, 4137 (2000). 

\bibitem{topological} S.~A.~Kivelson and V.~J.~Emery, Synth. Metals {\bf 80}, 
151 (1996).

\bibitem{chx} For a review, see A.~J.~Heeger, S.~A.~Kivelson, 
J.~R.~Schrieffer and W.~P.~Su, Rev. Mod. Phys. {\bf 60}, 781 (1988).

\bibitem{basov} Preliminary evidence of such behavior in very 
lightly doped LSCO has been recently observed by Dumm, Basov and Ando 
(private communication).

\bibitem{yamada} K.~Yamada, C.~H.~Lee, K.~Kurahashi, J.~Wada, S.~Wakimoto, 
S.~Ueki, H.~Kimura, Y.~Endoh, S.~Hosoya, G.~Shirane, R.~J.~Birgeneau, 
M.~Greven, M.~A.~Kastner and Y.~J.~Kim, Phys. Rev. B {\bf 57}, 6165 (1998).

\bibitem{mook} P.~Dai, H.~A.~Mook, R.~D.~Hunt and F.~Do\u{g}an, Phys. Rev. B
{\bf 63}, 54525 (2001).

\end{references}
\end{document}